# Modelling of Optimal Design of Manufacturing Cell Layout Considering Material Flow and Closeness Rating Factors

[1]Ghosh, T., [2]Dan, P., K.,

[1]Department of Industrial Engineering & Management, West Bengal University of Technology,
BF-142, Salt Lake City, PIN-700064
*tamal.31@gmail.com*
[2]Department of Industrial Engineering & Management, West Bengal University of Technology,
BF-142, Salt Lake City, PIN-700064
*danpk.wbut@gmail.com*

***Abstract:-*** Developing a group of machine cells and their corresponding part families to minimize the inter-cell and intra-cell material flow is the basic objective of the designing of a cellular manufacturing system (CMS). Afterwards achieving a competent cell layout is essential in order to minimize the total inter-cell part travels, which is principally noteworthy. There are plentiful articles of CMS literature which considered cell formation problems; however cell layout topic has rarely been addressed. Therefore this research is intended to focus on an adapted mathematical model of the layout design problem considering material handling cost and closeness ratings of manufacturing cells. Owing to the combinatorial class of the said problem, an efficient NP-hard technique based on Simulated Annealing metaheuristic is proposed henceforth. Some test problems are solved using the proposed technique. Computational results show that the proposed metaheuristic approach is extremely effective and efficient in terms of solution quality and computational complexity.

***Keywords:*** Cellular Layout, Inter-Cell Material Flow, Simulated Annealing, Closeness Ratings.

## 1. INTRODUCTION

Recently manufacturing and service farms are being operated more commendably with reduced throughput time because of intense competitive market and forced to invest less for all the important resources such as raw materials, assets and work forces to reduce the overall production costs or shorten the production lead time [1]. Among the latest manufacturing philosophies, group technology (GT) substantially reduces the throughput time and material flow in order to reduce the work in progress and finished goods inventories and enhances the forecast inaccuracies in dynamic production conditions [2]. Cellular manufacturing (CM) could be exemplified as a hybrid system associating the benefits of both the jobbing (flexibility) and mass (efficient flow and high production rate) production tactics which exploits the phenomenon of GT. The primitive concepts behind the designing of CM are, 1) to decompose the manufacturing system into cells by recognising and exploiting the similarities amongst components and machineries, 2) to design efficient inter-cell and intra-cell layout in order to smoothen the material flow on shop-floor. A huge number of articles are published in the domain of CM portraying various techniques to form efficient manufacturing cells [3], However layout problems are not addressed significantly. A competent layout in CMS not only improves its performance but also minimizes nearly 40-50% of the total production cost [4]. Cellular Layouts can be classified as inter-cell and intra-cell layouts. As soon as the cells are formed, those are believed to be assigned to optimal locations in order to minimize the inter-cell material flows, which is categorised as inter-cell layout problems [5]. Intra-cell layout is the optimal arrangement of machines inside the cell, which is also known as machine layout design. In reference [6] a mathematical model is proposed that considered the sequence of operations in evaluating the inter-cell and intra-cell moves and the impact of the cell layout to illustrate the inter-cell material flow. Sarker and Yu reported a twofold procedure for duplicating bottleneck machines in CMS environment in order to solve an inter-cell layout problem which assigns cells to locations to minimize the total inter-cell material flow [7]. Lee adopted an intra-cell and inter-cell layout design using three-phase interactive method which follows the decomposition strategy to reduce the large problem into a subset of smaller problems with minute details [8]. Wang and Sarker prescribed a 3-pair comparison heuristic and `bubble search' technique based inter-cell layout design algorithms in order to minimize the inter-cell material flow incurred due to bottleneck machines and also stated a lower bound on the QAP problem [9]. Solimanpur et al. developed an ant colony algorithm to solve the QAP model of inter-cell layout design problem and compared their results with other layout techniques such as H63, HC63-66, CRAFT and bubble search with improved solutions [10]. Kulkarni and Shankar employed a GA to the inter-cell layout problem and validated the performance of the algorithm with well-known layout design techniques [5]. Ariafar and Ismail proposed a new QAP model for inter-cell and intra-cell layout and solved using an SA algorithm with optimal solution [11]. Ma and Zhang demonstrated the dynamic layout framework based on reconfigurable CMS aiming at the enterprise problems in layout and production operation which is based on alternative process routing and multiple machine types available for operation considering cell formation and inter-cell layout jointly [12]. Jolai et al. employed a binary PSO based new heuristic approach to solve a QAP model for inter-cell and intra-cell layout problem considering parts demand and batch size using a variable neighborhood search [13]. Leno et al. recently discussed the multi-objective issues, a) minimization of material flow b) maximization of distance-weighted closeness factor of cells and solved that using a GA with near optimal solutions [14].

In this article, a mathematical model is developed depicted in next section. In section 3 the proposed Simulated Annealing technique is elaborated. Section 4 demonstrates the results and discussion. Section 5 discusses concludes this work.

## 2. PROBLEM FORMULATION

The layout problem in CMS is a Quadratic Assignment Problem (QAP) which restrains each cell to be assigned to only one location and each location to be selected for only one cell. Mathematically the distance or the unit cost of travel between two locations $j$ and $l$ is expressed as $d_{jl}$ (j, l = 1, 2, 3…, N). The inter-cell material flow is expressed as $f_{ik}$ (i, k = 1, 2, 3, …, N) is the amount of material flow from cell $i$ to $k$. Traditional single period cell layout problems could be formulated with the help of following QAP mathematical model [5],

$$Minimize\ Z = \sum_{i=1}^{N}\sum_{j=1}^{N}\sum_{k=1}^{N}\sum_{l=1}^{N} f_{ik}d_{jl}x_{ij}x_{kl} \quad (1)$$

Subject to,

$$\sum_{i=1}^{N} x_{ij} = 1 \quad\quad j \in N \quad\quad (2)$$

$$\sum_{j=1}^{N} x_{ij} = 1 \quad\quad i \in N \quad\quad (3)$$

$$x_{ij} = \begin{cases} 1, & if\ cell\ i\ is\ assigned\ to\ location\ j \\ 0, & Otherwise \end{cases} \quad (4)$$

Constraint (2) restrains the assignment of one cell at one location, and constraint (3) makes sure that each location can only be designated to only one cell. $x_{ij}$ is the decision variable.

In the domain of CMS, past literature rarely presents any multi-objective cellular layout model which practically formulated quantitative and qualitative objectives simultaneously. Only reference [14] proposed a multi-objective model for SPCLP recently. Closeness rating relationships are commonly used in layout design literature [15, 16]. This relationship could be defined as,

$r_{ik}$ = closeness rating between cells $i$ and $k$.

This fact could be realised using some suitable example. For that matter an example dataset is considered in the Table 1. Since this relationship is qualitative in nature, therefore it reflects qualitative values of the relationship and illustrated using scoring system.

For example, the adjacency relationships can be demonstrated by the following numerical values, *A = 6, E = 5, I = 4, O = 3, U = 2 and X = 1.*

Table 1. Closeness relationship matrix of the cells (6×6 dataset)

|    | C1 | C2 | C3 | C4 | C5 | C6 |
|----|----|----|----|----|----|----|
| C1 | -  | E  | O  | U  | A  | I  |
| C2 |    | -  | E  | U  | A  | U  |
| C3 |    |    | -  | X  | U  | X  |
| C4 |    |    |    | -  | U  | U  |
| C5 |    |    |    |    | -  | A  |
| C6 |    |    |    |    |    | -  |

Thus the 2nd objective function becomes,

$$Maximize\ Z2 = \sum_{i=1}^{N}\sum_{j=1}^{N}\sum_{k=1}^{N}\sum_{l=1}^{N} r_{ikp}d_{jl}x_{ij}x_{kl} \quad (5)$$

As soon as all the relationship factors are quantified, a method of normalisation of data is applied using the following formula adopted from Harmonosky and Tothero [15],

$$N_{ik} = \frac{S_{ik}}{\sum_{i=1}^{N}\sum_{k=1}^{N} S_{ik}} \quad (6)$$

Where,

$S_{ikf}$ is the relationship value between cells $i$ and $k$.

$N_{ikf}$ is the normalized relationship value between cells $i$ and $k$.

Therefore the final weighted QAP formulation is the combined form of $Z1$ and $Z2$,

$$Minimize\ Z = \sum_{i=1}^{N}\sum_{j=1}^{N}\sum_{k=1}^{N}\sum_{l=1}^{N} Nf_{ik} d_{jl} x_{ij} x_{kl} + w \times \sum_{i=1}^{N}\sum_{j=1}^{N}\sum_{k=1}^{N}\sum_{l=1}^{N} Nr_{ik} d_{jl} x_{ij} x_{kl} \quad (7)$$

Subject to,

$$\sum_{i=1}^{N} x_{ij} = 1 \quad j \in N \quad (8)$$

$$\sum_{j=1}^{N} x_{ij} = 1 \quad i \in N \quad (9)$$

$$x_{ij} = \begin{cases} 1, & \text{if cell } i \text{ is assigned to location } j \\ 0, & \text{Otherwise} \end{cases} \quad (10)$$

Equation (7) is the multi-objective QAP formulation which is a minimization type function. $Nf_{ik}$ and $Nr_{ik}$ are the normalized value of $f_{ik}$ and $r_{ik}$ respectively. Equation (8) and (9) are the assignment constraints, ensures that each location contains only one cell and each cell is assigned to only one location. Remaining constraint is the relationship of decision variable $x_{ij}$. In the objective function '$w$' is the load factor of the qualitative part $Z2$, which is adopted from Urban [17]. In his article, Urban suggested to select the value of '$w$' as the largest value of the material flow matrix in order to transform the qualitative term to correspond with the quantitative volumes. However Urban didn't utilize the normalization of the quantitative or qualitative term as it is done by Harmonosky and Tothero [16]. Due to the adoption of that phenomenon, both the qualitative and quantitative objectives are transformed into same scale. Thus the value of the constant '$w$' is fixed in the range $0 \leq w \leq 1$. If '$w$' takes the value 0, it means the cost function transformed purely into material handling cost and if it takes value 1, then both material flow and relative proximity factor would share same importance. However in experimental stage the QAP model of cost function is tested for four values of '$w$', which are 0.2, 0.4, 0.6, and 0.8.